\documentclass[10pt,twocolumn,journal]{IEEEtran}

\usepackage{amsmath, mathrsfs}
\usepackage{stmaryrd}
\usepackage{relsize}

\usepackage{cite}
\usepackage{amsfonts}
\usepackage{amssymb}
\usepackage{color}
\usepackage{multirow}
\usepackage{graphicx}
\usepackage{relsize}
\usepackage{multirow}
\usepackage{multicol}
\usepackage{caption}
\usepackage{subcaption}
\usepackage[numbers,sort&compress]{natbib}
\usepackage{balance}
\usepackage{arydshln}

\allowdisplaybreaks

\usepackage{amsfonts}
\usepackage{amssymb}
\usepackage{graphicx}
\usepackage{color}
\usepackage{multirow}
\usepackage{graphicx}

\usepackage{caption}
\usepackage{subcaption}
\usepackage[numbers,sort&compress]{natbib}

\newcommand{\subparagraph}{}
\usepackage[compact]{titlesec}
\titlespacing*{\section}{15pt}{1\baselineskip}{0.9\baselineskip}

\allowdisplaybreaks

\newcommand{\myhash}{%
  {\settoheight{\dimen0}{C}\kern-.05em\, \resizebox{!}{\dimen0}{\raisebox{\depth}{\#}}}}

\usepackage{multicol}


\usepackage{mathbbol}

\def\mindex#1{\index{#1}}

\def\sq{\hbox{\rlap{$\sqcap$}$\sqcup$}}
\def\qed{\ifmmode\sq\else{\unskip\nobreak\hfil
\penalty50\hskip1em\null\nobreak\hfil\sq
\parfillskip=0pt\finalhyphendemerits=0\endgraf}\fi\medskip}

\long\def\defbox#1{\framebox[.9\hsize][c]{\parbox{.85\hsize}{%
\parindent=0pt
\baselineskip=12pt plus .1pt      
\parskip=6pt plus 1.5pt minus 1pt 
 #1}}}

\long\def\beginbox#1\endbox{\subsection*{}%
\hbox{\hspace{.05\hsize}\defbox{\medskip#1\bigskip}}%
\subsection*{}}

\def\endbox{}

\def\diag{{\text{diag}}}

\newsavebox{\junk}
\savebox{\junk}[1.6mm]{\hbox{$|\!|\!|$}}

\def\bC{{\mathbb C}}

\def\bfB{{\bf B}}

\def\bfH{{\bf H}}
\def\bfI{{\bf I}}

\def\bfT{{\bf T}}

\def\bfW{{\bf W}}
\def\bfX{{\bf X}}

\def\bfa{{\bf a}}

\def\bfh{{\bf h}}

\def\bfn{{\bf n}}

\def\bfr{{\bf r}}
\def\bfs{{\bf s}}

\def\bfx{{\bf x}}
\def\bfy{{\bf y}}

\def\sfH{{\sf H}}

\def\bfmath#1{{\mathchoice{\mbox{\boldmath$#1$}}%
{\mbox{\boldmath$#1$}}%
{\mbox{\boldmath$\scriptstyle#1$}}%
{\mbox{\boldmath$\scriptscriptstyle#1$}}}}

\def\bfmY{\bfmath{Y}}

\def\bfmhhaY{\bfmath{\hhaY}} 
\def\bfmhhaY{\hbox to 0pt{$\widehat{\bfmY}$\hss}\widehat{\phantom{\raise 1.25pt\hbox{$\bfmY$}}}}

\def\til={{\widetilde =}}

\def\clC{{\cal C}}

\def\clN{{\cal N}}

 \def\FRAC#1#2#3{\genfrac{}{}{}{#1}{#2}{#3}}

\def\ddtp{{\mathchoice{\FRAC{1}{d^{\hbox to 2pt{\rm\tiny +\hss}}}{dt}}%
{\FRAC{1}{d^{\hbox to 2pt{\rm\tiny +\hss}}}{dt}}%
{\FRAC{3}{d^{\hbox to 2pt{\rm\tiny +\hss}}}{dt}}%
{\FRAC{3}{d^{\hbox to 2pt{\rm\tiny +\hss}}}{dt}}}}

\def\average#1,#2,{{1\over #2} \sum_{#1}^{#2}}

\def\eye(#1){{\bf(#1)}\quad}

\def\eq#1/{(\ref{e:#1})}

\newcommand{\beqn}[1]{\notes{#1}%
\begin{eqnarray} \elabel{#1}}

\newcommand{\eeqn}{\end{eqnarray} }

\newcommand{\beq}[1]{\notes{#1}%
\begin{equation}\elabel{#1}}

\newcommand{\eeq}{\end{equation}}

\def\bdes{\begin{description}}
\def\edes{\end{description}}

\newcounter{rmnum}

\newcounter{anum}

{\end{list}}

\def\ass(#1:#2){(#1\ref{#1:#2})}

\def\ritem#1{
\item[{\sf \ass(\current_model:#1)}]
}

\newenvironment{recall-ass}[1]{%
\begin{description}
\def\current_model{#1}}{
\end{description}
}

\usepackage{tikz}
\usepackage{pgfplots,tikz-3dplot}
\pgfplotsset{compat=newest}
\usetikzlibrary{patterns}
\usetikzlibrary{positioning}
\usetikzlibrary{datavisualization}
\usetikzlibrary{datavisualization.formats.functions}
\usetikzlibrary{backgrounds}
\usetikzlibrary{shapes,snakes}

\usepgfplotslibrary{external} 
\usepackage{float}
\usepackage{afterpage}
\usepackage{balance}

\usepackage{geometry}
\newcommand{\xiul}{{\mathcal{I}}_{{\rm ul},i}}
\newcommand{\xidl}{{\mathcal{I}}_{{\rm dl},i}}
\newcommand{\Xgam}{\mathcal{X}_{\gamma}}
\newcommand{\Xgamhat}{\hat{\mathcal{X}}_{\gamma}}

\newcommand{\aulvec}{\bfa_{\rm ul}(\theta)}

\newcommand{\adlvec}{\bfa_{\rm dl}(\theta)}
\newcommand{\Pbf}{P}

\newcommand{\Bcup}{\mathlarger{\mathlarger{\cup}}}

\newcommand{\suppul}{{\cal S}_{\rm ul}}
\newcommand{\suppulhat}{\hat{{\cal S}}_{\rm ul}}

\newcommand{\suppdl}{{\cal S}_{\rm dl}}
\newcommand{\suppdlhat}{\hat{{\cal S}}_{\rm dl}}

\newcommand{\htul}{\bfh_{\rm ul}}
\newcommand{\hul}{\check{\bfh}_{\rm ul}}

\newcommand{\htdl}{{\bfh}_{\rm dl}}
\newcommand{\hdl}{\check{\bfh}_{\rm dl}}

\newcommand{\heff}{{\bfh}_{\rm eff}}

\let\emptyset\varnothing
\def\herm{{\sfH}}

\def\cg{{\clC\clN}}

\long\def\comment#1{}

\newfont{\bb}{msbm10 scaled 1100}

\newcommand{\EE}{\mbox{\bb E}}

\newcommand{\hv}{{\bf h}}

\newcommand{\nv}{{\bf n}}

\newcommand{\vv}{{\bf v}}

\newcommand{\yv}{{\bf y}}

\newcommand{\Fm}{{\bf F}}

\newcommand{\Id}{{\bf I}}

\newcommand{\Nm}{{\bf N}}

\newcommand{\Qm}{{\bf Q}}

\newcommand{\Xm}{{\bf X}}
\newcommand{\Ym}{{\bf Y}}

\newcommand{\Ac}{{\cal A}}
\newcommand{\Bc}{{\cal B}}

\newcommand{\Ec}{{\cal E}}

\newcommand{\Gc}{{\cal G}}

\newcommand{\Ic}{{\cal I}}

\newcommand{\Kc}{{\cal K}}

\newcommand{\Oc}{{\cal O}}

\newcommand{\Sc}{{\cal S}}

\renewcommand{\arg}{{\hbox{arg}}}

\newcommand{\SNR}{{\sf SNR}}

\newcommand{\transp}{{\sf T}}

\newcommand{\aul}{\bfa_{\rm ul}(\theta)}
\newcommand{\adl}{\bfa_{\rm dl}(\theta)}

\graphicspath{ {./Figures/} }

\begin{document}

\title{FDD Massive MIMO: Efficient Downlink Probing and Uplink Feedback via Active Channel Sparsification}
\author{\IEEEauthorblockN{Mahdi Barzegar Khalilsarai\IEEEauthorrefmark{1},
Saeid Haghighatshoar\IEEEauthorrefmark{1}, Xinping Yi\IEEEauthorrefmark{2}, Giuseppe Caire\IEEEauthorrefmark{1}}\\
\IEEEauthorblockA{\IEEEauthorrefmark{1}\IEEEauthorblockA{Communications and Information Theory Group, Technische Universit\"{a}t Berlin}}\\
\IEEEauthorblockA{\IEEEauthorrefmark{2}\IEEEauthorblockA{Department of Electrical Engineering and Electronics, University of Liverpool}}\\
Emails: $^*\{$m.barzegarkhalilsarai, saeid.haghighatshoar, caire$\}$@tu-berlin.de, $^\dagger$xinping.yi@liverpool.ac.uk}

\maketitle

\begin{abstract}
In this paper, we propose a novel method for efficient implementation of a massive Multiple-Input Multiple-Output (massive MIMO) system with Frequency Division Duplexing (FDD) operation. Our main objective is to reduce the large overhead incurred by Downlink (DL) common training and Uplink (UL) feedback needed to obtain channel state information (CSI) at the base station. Our proposed scheme relies on the fact that the underlying angular distribution of a channel vector, also known as the angular scattering function, is a frequency-invariant entity yielding a UL-DL reciprocity and has a limited angular support. We estimate this support from UL CSI and interpolate it to obtain the corresponding angular support of the DL channel. Finally we exploit the estimated support of the DL channel of all the users to design an efficient channel probing and feedback scheme that maximizes the total spectral efficiency of the system.  Our method is different from the existing compressed-sensing (CS) based techniques in the literature. Using support information helps reduce the feedback overhead from $\Oc(s \log M)$ in CS techniques to $\Oc (s)$ in our proposed method, with $s$ and $M$ being sparsity order of the channel vectors and the number of base station antennas, respectively. Furthermore, in order to control the channel sparsity and therefore the DL common training and UL feedback overhead, we introduce the novel concept of \textit{active channel sparsification}. In brief, when the fixed pilot dimension is less than the required amount for reliable channel estimation, we introduce a pre-beamforming matrix that artificially reduces the effective channel dimension of each user to be not larger than the DL pilot dimension, while maximizing both the number of served users and the number of probed angles. We provide numerical experiments to assess the performance of our method and compare it with the state-of-the-art CS technique.
\end{abstract}


\section{Introduction}
Massive base station (BS) antenna arrays (massive MIMO) promise huge improvements in a variety of aspects, including data rate, reliability, energy efficiency and interference reduction in wireless networks \cite{larsson2014massive}. Realizing massive MIMO with Time Division Duplexing (TDD) operation is convenient, due to the inherent Uplink-Downlink (UL-DL) channel reciprocity \cite{marzetta2006much}. In contrast, channel reciprocity does not hold in Frequency Division Duplexing (FDD) operation, since UL and DL take places in different bands, which are separated by much more than the fading coherence bandwidth. Therefore, the UL channel state information (CSI) can not be used for DL data transmission, so that the BS has to probe the DL channel via training and ask for CSI feedback from the users. Both the DL training and UL feedback impose huge overheads, particularly in massive MIMO systems. For example, by conventional orthogonal training the BS needs $T\ge M$ pilot symbols to train $M$ antennas and since $M\gg 1$, the BS may lack enough signal dimensions even for training the channel. In addition, we have a similar problem in UL, where the users have to feedback the high-dimensional CSI, which will consume a large part of the available UL signal dimensions. 

 Despite these issues, FDD massive MIMO systems are still desirable because most of the current wireless networks are based on FDD and FDD systems show a better performance in scenarios with symmetric traffic and delay-sensitive applications \cite{jiang2015achievable,rao2014distributed}. In recent years numerous techniques have been proposed to reduce the DL training and UL feedback overhead in FDD massive MIMO systems. Some of these techniques rely on code-book based CSI quantization and fall in two categories: the designs based on time correlation of the channel vectors \cite{choi2015trellis,heath2009progressive} and the designs based on spatial correlation of the channel vectors \cite{jiang2015achievable,adhikary2013joint,adhikary2014joint,nam2014fundamental}. Other techniques are based on exploiting the low-rank or sparse structures to reduce DL training overhead, since in a massive MIMO scenario, the received signal from a user at the BS consists of a few multi-path components with a limited Angle of Arrival (AoA) support, resulting in a sparse representation. Compressed sensing (CS) methods leverage this structure to recover the channel vector at the user side, from a handful of measurements received during DL channel probing. An important example of these methods is presented in \cite{rao2014distributed}. In this work, the user channels are estimated via running a Joint Orthogonal Matching Pursuit (J-OMP) algorithm on the compressed channel measurements collected from all users, achieving a considerable reduction in the feedback overhead. In \cite{kuo2012compressive}, compressed channel feedback methods for spatially correlated channels are proposed by introducing a sparsifying dictionary for the channel vector based on Karhunen-Lo\`eve transform (KLT). A dictionary-learning based approach for sparse channel modeling is presented in \cite{ding2016dictionary}. Exploiting angular UL-DL reciprocity, this work proposes a joint UL-DL sparsifying dictionary which allows for compressed channel estimation with much fewer measurements.
 
In this paper we focus on the UL-DL angular reciprocity, which is characterized in terms of a continuous, frequency-invariant angular scattering function, modeling the density of the power received from the user in the AoA domain. We make the key observation that although the channel vectors in the UL and DL are statistically independent from each other, thus, the channel reciprocity in the traditional sense does not hold, we still have a type of reciprocity due to the fact that the angular scattering function is the same for UL and DL transmission. We refer to this feature as \textit{the reciprocity of the angular scattering function}. This is a manifestation of the angular reciprocity which is already known and exploited in the literature \cite{vasisht2016eliminating,Xie2017Channel,ding2016dictionary}, expressed in more general terms since we do not assume necessarily discrete, seperable angles, but a continuum of AoAs. This is important because explicit superresolution angle estimation (such as the one presented in \cite{vasisht2016eliminating}) would fail in the presence of a continuous angular scattering function, while our method works regardless of the shape of the angular scattering function. We exploit the reciprocity of this function to derive the DL channel support given UL channel observations. This information helps us to estimate the DL channel using far fewer measurements even compared to CS based methods, in the order of the maximum angular sparsity of DL channel vectors. This results in a huge reduction in the necessary DL training dimension and UL feedback overhead.

In general, when the number of DL pilot dimensions is less than the channel sparsity (number of significant coefficients in the angular basis with respect to which the channel is sparse), any estimation technique yields a very large estimation error. Here, we also propose a method to partially estimate all DL channel vectors even with a very small pilot dimension. This method is referred to as \textit{active channel sparsification} and is obtained by introducing a pre-beamforming (pre-BF) matrix, which can be optimized such that the effective channel dimension of any user is not larger than the DL pilot dimension, so that all effective channels can be estimated, while the overall number of probed (and eventually served) users and signal space dimension are maximized. Active channel sparsification is done using a linear integer programming (ILP) optimization problem, which can be solved using off-the-shelf solvers such as MATLAB.
%

We denote vectors by boldface small letters (e.g. $\bfx$), matrices by boldface capital letters (e.g. $\bfX$), scalars by non-boldface letters (e.g. $x$ or $X$), and sets by calligraphic letters (e.g. $\mathcal{X}$). The $i$\textsuperscript{th} element of a vector $\bfx$ and the $(i,j)$\textsuperscript{th} element of a matrix $\bfX$ will be denoted by $[{\bfx}]_i$ and $[{\bfX}]_{i,j}$. For a matrix $\bfX$, we denote its $i$\textsuperscript{th} row and $j$\textsuperscript{th} column with the row vector $\bfX_{i,.}$ and the column vector $\bfX_{.,j}$, respectively. We use the shorthand notation $[k]$ to denote the set of integers $\{1, . . . , k\}$. For arguments that are intervals over the real line, $|\cdot|$ returns the length of the interval and for arguments that are discrete sets, it returns the cardinality of the set. We always denote the identity matrix of order $p$ with $\bfI_p$. 
\begin{figure}[t]
	\centering
	\includegraphics[width=0.45\textwidth]{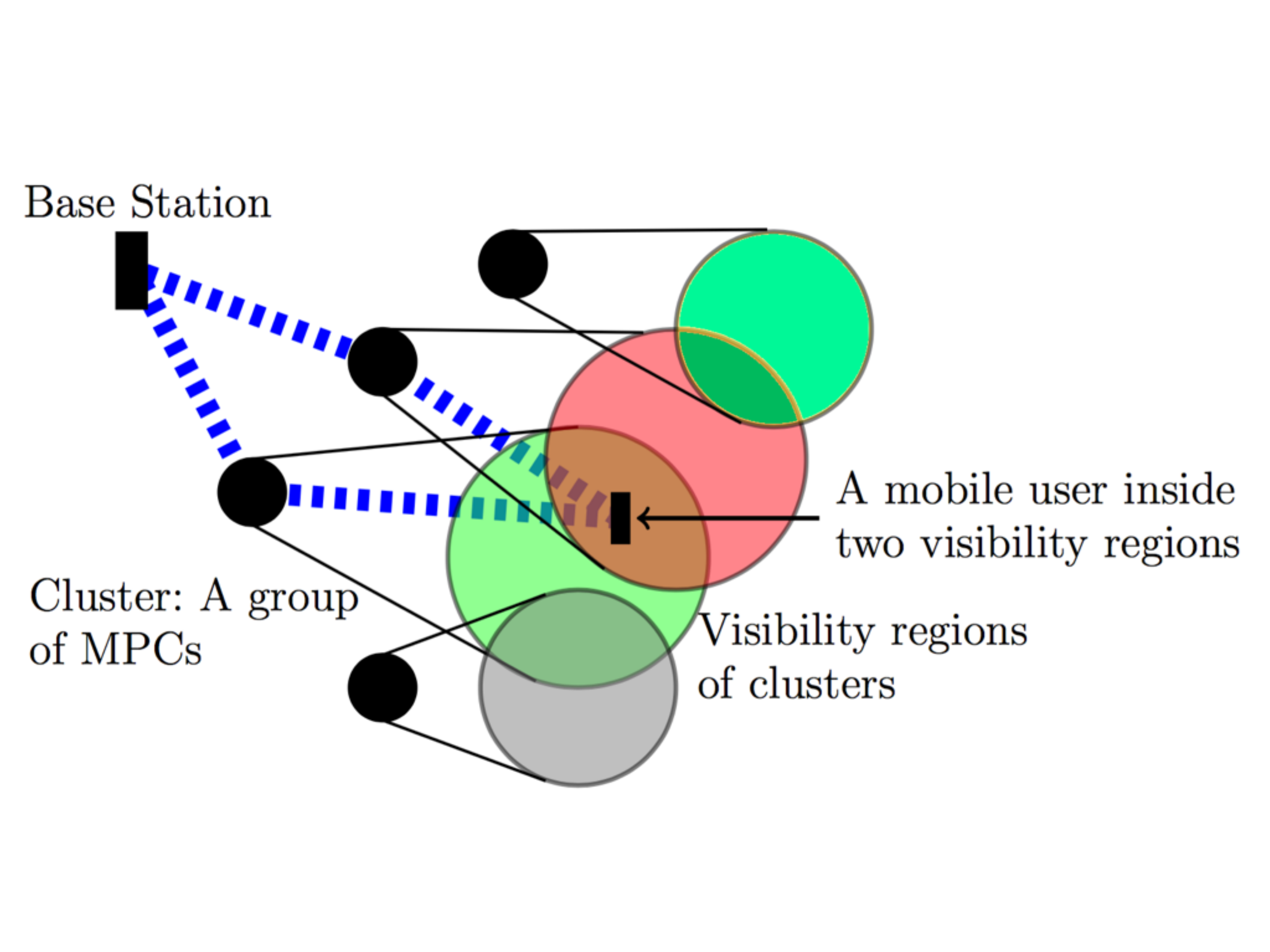}
	\caption{A sketch of MPC clusters and visibility regions in a propagation environment.}
	\label{fig:Cost}
\end{figure}
\section{System Setup}\label{sec:2}

We consider the geometry-based stochastic channel model (GSCM), which consists of clusters of multipath components (MPCs) and visibility regions (VRs) as its building blocks. An MPC cluster is generated by the reflection of signal from objects in the environment. A visibility region represents the region over which the signal from a user can reach the BS by propagating through a particular MPC cluster. Fig. \ref{fig:Cost} illustrates a sketch of the described model. By adopting GSCM one can assume that the channel scattering geometry for a user is piecewise time invariant, since moving across a VR occurs in time scales much larger than moving across one wave-length. Hence we focus on this piecewise stationary situation and consider the channel model for a fixed scattering geometry.

\begin{figure}[t]
	\centering
	\begin{subfigure}[b]{0.5\textwidth}
		\centering
		\includegraphics[width=0.7\textwidth]{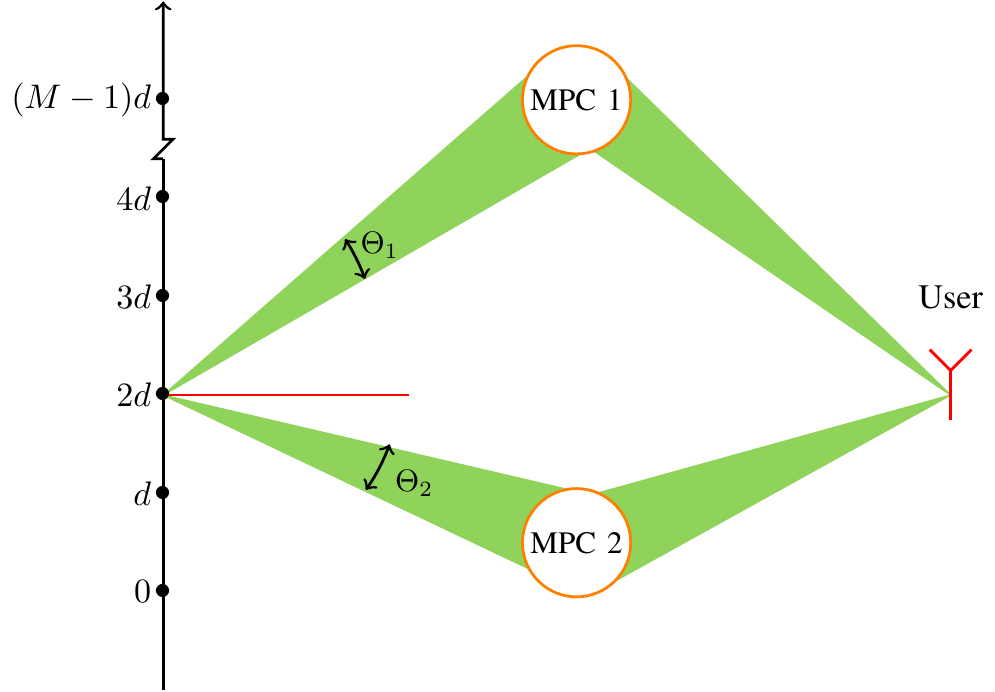}
		\caption{}
		\label{ang_intervals}
	\end{subfigure}
	
	\begin{subfigure}[b]{0.5\textwidth}
		\centering
		\includegraphics[width=0.7\textwidth]{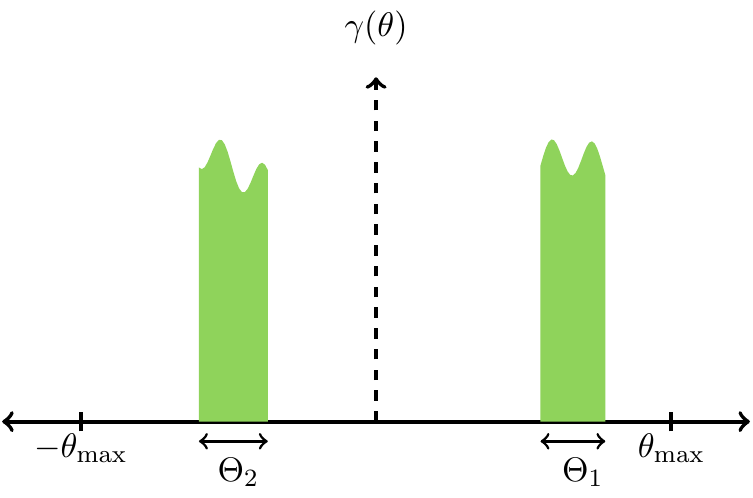}
		\caption{}
		\label{example_scat_func}
	\end{subfigure}	
	\caption{(a) An example of the propagation geometry. (b) The angular scattering function corresponding to the example geometry in (a).}
\end{figure}

Consider a BS equipped with a uniform linear array (ULA) of $M\gg 1$ antennas and a user with a single antenna. Fig. \ref{ang_intervals} illustrates an example of the propagation geometry for a single user along with the array formation. During UL, the signal is received at the BS through a continuum of AoAs and for a time-frequency resource block it can be written as $\bfr = \check{\bfh}_{\rm ul} x + \bfn$, where 
\begin{equation}\label{h_ul}
\check{\bfh}_{\rm ul}:=\int_{\Theta} \rho_{\rm ul}(\theta) \aulvec {\rm d}\theta \in \bC^M
\end{equation}
denotes the UL channel vector, where $\Theta:=[-\theta_{\max}, \theta_{\max})$ is the angular range scanned by the BS array, where $x\in \bC$ is the transmitted UL pilot symbol of the user along the channel vector $\check{\bfh}_{\rm ul}$, which typically belongs to a signal constellation such as QAM, where $\bfn \sim \cg({\bf 0}, \sigma^2\Id_M)$ is the Additive White Gaussian Noise (AWGN) of the  antenna elements, and where $\aulvec \in \bC^M$ is the UL array response at AoA $\theta$, whose $\ell$\textsuperscript{th} component is given by $[\bfa_{\rm ul}(\theta) ]_\ell= e^{j \frac{2\pi}{c} f_{\rm ul} \ell d \sin\theta}, $
where $f_{\rm ul} $, $c$ and $d$ are the carrier frequency over the UL band, the speed of light, and the antenna spacing, respectively. In \eqref{h_ul}, $\rho_{\rm ul}(\theta) $ denotes a complex, circularly symmetric, zero-mean, Gaussian random process representing the random gain of the scatterers at different AoAs. This random process is completely characterized by its second order statistics. 
\begin{equation} 
\EE[\rho_{\rm ul}(\theta) \rho_{\rm ul}^*(\theta')]  = \gamma(\theta) \delta(\theta - \theta'),
\end{equation}
where $\gamma(\theta)$ is the angular scattering function, which represents the received signal energy density as a function of the AoA. Fig. \ref{example_scat_func} illustrates the angular scattering function corresponding to the geometry presented in Fig. \ref{ang_intervals}. Since the MPC clusters occupy only a limited portion of the angular range, $\gamma (\theta)$ has a limited support, denoted by $\Xgam:= \{\theta : \gamma(\theta) \neq 0\}$. As a result, as we will show, the channel vectors generated by this angular scattering function are approximately sparse in the Fourier basis. Let $\htul := \Fm^\herm\hul $ be the vector of Fourier coefficients for $\hul$, where $\Fm \in \bC^{M\times M}$ is the DFT matrix whose $(k,\ell)$ element is given by $\left[\Fm\right]_{k,\ell} := 	\tfrac{1}{\sqrt{M}} {\rm e}^{{\rm j} \frac{2\pi}{M}k (\ell -\frac{M}{2})}$.
We show that only a few entries in $\htul$ have a significant variance. First note that $\EE\{\htul\} = \int_{\Theta} \EE\left\{\rho_{\rm ul}(\theta)\right\} \Fm^\herm  \aul {\rm d}\theta = \bf0$. The variance of each component in $\htul$ is given by the corresponding component in the vector $\vv_{\rm ul}=\diag\left( \int_{\Theta} \gamma (\theta) \Fm^\herm \aul  \aul^\herm \Fm {\rm d} \theta \right)$. By a simple calculation one can show that

%

\begin{equation}\label{eq:var_vec}
\left[\vv_{\rm ul}\right]_i = \frac{1}{M}  \int_{\Theta}   \gamma(\theta)    \left|       
\frac{\sin \left ( \pi \psi_{{\rm ul},i}(\theta) M \right )}{\sin \left ( \pi \psi_{{\rm ul},i} (\theta)\right )} \right|^2  {\rm d}\theta,
\end{equation}
	where we have defined $\psi_{{\rm ul},i}(\theta) := \frac{d}{c} f_{\rm ul}\sin \theta - \frac{i}{M}+\frac{1}{2}$. The function $D_M(\psi) := \frac{\sin(\pi \psi M)}{\sin (\pi \psi)}$ is the Dirichlet kernel with parameter $M$ and as we know $\left| D_M(\psi) \right|^2$ has a significant magnitude only for $|\psi| \le 1/M$, hence $\left|       
	\frac{\sin \left ( \pi \psi_{{\rm ul},i}(\theta) M \right )}{\sin \left ( \pi \psi_{{\rm ul},i}(\theta) \right )} \right|^2$ is non-negligible only for those angles $\theta$ for which $|\psi_{{\rm ul},i}(\theta)| \leq \frac{1}{M}$, i.e. for $\theta \in \xiul$, where
	$\xiul	=\{\theta : |\psi_{{\rm ul},i}(\theta)| \leq \frac{1}{M} \}$. From this observation and \eqref{eq:var_vec} we conclude that the $i$\textsuperscript{th} element in $\htul$ has significant variance if and only if $\xiul \cap \Xgam \neq \emptyset$ and since $|\xiul| \approx \Oc (1/M)$, the vector $\htul$ has significant variance only for a small set of indices $i$. We denote this set by the support set $\suppul$ and define it as
	\begin{equation}
	\suppul = \{ i\in [M] : \xiul \cap \Xgam \neq \emptyset \}.
	\end{equation} 
Furthermore, since the support of $\gamma (\theta)$ consists of intervals over the real line corresponding to MPC clusters, we expect $\htul$ to have a block-sparse structure. In other words, the non-zero elements in $\htul$ come in clusters.

The DL channel vector can be described in a similar way by $\hdl :=\int_{\Theta} \rho_{\rm dl}(\theta) \adlvec {\rm d}\theta \in \bC^M$, where $[\adlvec]_\ell= e^{j \frac{2\pi}{c} \ell d f_{\rm dl} \sin\theta}$ denotes the array response in DL with $f_{\rm dl}$ being the DL carrier frequency. Note that the generating Gaussian process of the DL channel, i.e. $\rho_{\rm dl}(\theta)$, has the same statistics as its UL counterpart $\rho_{\rm ul}(\theta)$. Defining the vector of Fourier coefficients for the DL channel vector by $\htdl := \Fm^\herm\hdl $ we have that $\EE\{\htdl\} = \int_{\Theta} \EE\left\{\rho_{\rm dl}(\theta)\right\} \Fm^\herm  \adl {\rm d}\theta = \bf0$ and the vector of variances is given by $\vv_{\rm dl}=\diag\left( \int_{\Theta} \gamma (\theta) \Fm^\herm \adl  \adl^\herm \Fm {\rm d} \theta \right)$ and $\left[\vv_{\rm dl}\right]_i = \frac{1}{M}  \int_{\Theta}   \gamma(\theta)    \left|       
\frac{\sin \left ( \pi \psi_{{\rm dl},i}(\theta) M \right )}{\sin \left ( \pi \psi_{{\rm dl},i} (\theta)\right )} \right|^2  {\rm d}\theta$. Here $\psi_{{\rm dl},i}(\theta) := \frac{d}{c} f_{\rm dl}\sin \theta - \frac{i}{M}+\frac{1}{2}$ with the only difference being the different carrier frequency in DL. In a similar fashion the support set in DL is given by 
	\begin{equation}
\suppdl = \{ i\in [M] : \xidl \cap \Xgam \neq \emptyset \},
\end{equation} 
where $\xidl=\{\theta : |\psi_{{\rm dl},i}(\theta)| \leq \frac{1}{M} \}$. With the same reasoning as before, $\htdl$ has a block-sparse structure. Having an estimate of the DL support helps the BS to probe the DL channel with much fewer measurements. Hence, our idea is to estimate the DL support set $\suppdl$ for each user using its UL pilot signals as described in the next section.

\section{Downnlink Support Estimation}\label{sec:3}
We assume that the channel is approximately constant across a resource block of $N_c=\Delta f_c \times \Delta t_c$ time-frequency tiles. We call each tile a signal dimension. The BS or the users devote a part of these $N_c$ signal dimensions to channel probing via pilot transmission and the remaining for data communication. Now, let assume that during UL each user sends $L$ pilot symbols $\{x_i\}_{i=1}^L$ through $L$ signal dimensions to the BS. Without loss of generality we can assume $x_i=1$ for all $i$. The received signal at the BS can be written as
\begin{equation}
\yv_i = \,  \check{\hv}_{{\rm ul},i} + \nv_i,
\end{equation}
where $\check{\hv}_{{\rm ul},i}$ is the channel vector corresponding to the $i$\textsuperscript{th} signal dimension and $\nv_i\sim \cg({\bf 0}, \sigma^2\Id_M)$ is the AWGN. We can safely assume that the vectors $\{\check{\hv}_{{\rm ul},i}\}_{i=1}^L$ share the same support set since the sparsity pattern depends only on the slow varying geometry of the propagation environment. There are plenty of denoising techniques to estimate the support set from noisy observations $\{\yv_i\}_{i=1}^L$, among which we choose the one presented in \cite{haghighatshoar2017massive}.  Define $\Ym = \left[\yv_1, \ldots, 	\yv_L\right]$ and $\Nm=\left[\nv_1,\ldots, \nv_L\right]$. The support estimation problem amounts to finding the set of indices corresponding to the non-zero rows of the solution matrix $\Xm^\ast \in \bC^{M\times L}$ in a \text{Multiple Measurement Vectors} (MMV) problem. This problem can be formulated as follows,
\begin{equation}\label{l2_1_first_formula}
\Xm^\ast\, =  ~\underset{\Xm \in \bC^{M\times L}}{\arg \min}~ \Vert \Xm\Vert_{2,1}, ~ \text{subject to} ~ \Vert \Ym - \Fm \Xm\Vert \le \sqrt{ML} ~\sigma, 
\end{equation}
where the $\ell_{2,1}$-norm is defined by $\Vert \Xm \Vert_{2,1}  = \sum_{i=0}^{L-1} \Vert \Xm_{i,.}\Vert_2$ and $\Vert \cdot \Vert $ denotes Frobenius norm. Once \eqref{l2_1_first_formula} is solved, we obtain the UL support set by calculating the $\ell_2$-norm of each row of matrix $\Xm$. If the $\ell_2$-norm of a particular row is greater than a certain threshold $\epsilon$, then it is labeled as \textit{active} and otherwise it is labeled as \textit{inactive}. In other words
\begin{equation}
\hat{\Sc}_{\rm ul} := \{i\in [\, M] : \Vert \Xm^\ast_{i,.}\Vert_2 \ge \epsilon \}
\end{equation}
denotes the estimated UL channel support of the user. 

\begin{figure}[t]
	\centering
	\begin{subfigure}[b]{0.5\textwidth}
		\centering
		\includegraphics[width=0.75\textwidth]{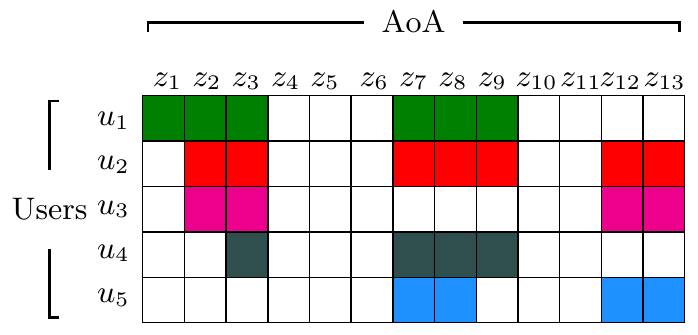}
		\caption{}
	\label{supp_sets}
	\end{subfigure}
	
	\begin{subfigure}[b]{0.5\textwidth}
		\centering
		\includegraphics[width=0.7\textwidth]{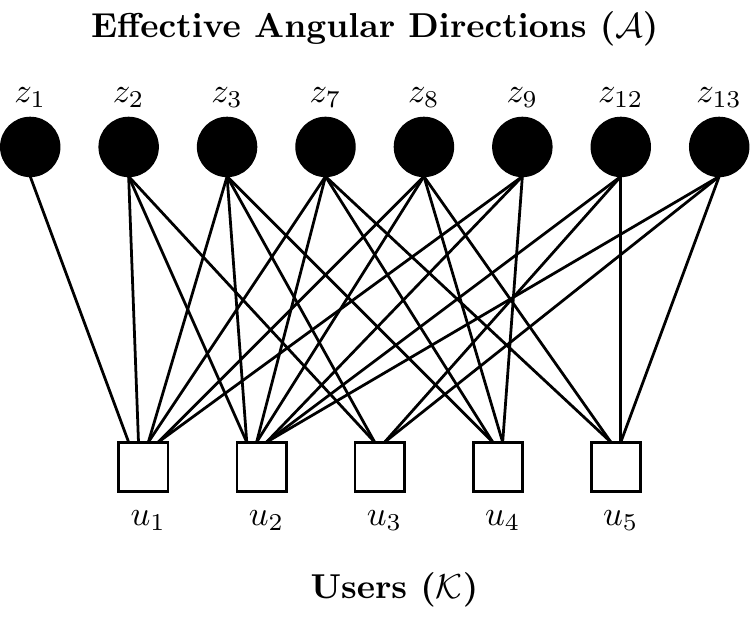}
		\caption{}
		\label{bi_graph}
	\end{subfigure}	
	\caption{(a) Schematic of estimated DL support profile for different users, available at the BS before DL probing. The colored grid points represent support elements. (b) The corresponding bipartite graph $\Gc$.}
\end{figure}

Now recall that each index $i$ in $\hat{\Sc}_{\rm ul}$ corresponds to an interval $\Ic_{{\rm ul}, i}$. As we described in the previous section, the variance of element $j$ is significant only if $ \xiul \cap \Xgam \neq \emptyset$. This gives us a hint that, if $i$ is in the support set, its corresponding interval $\Ic_{{\rm ul}, i}$ should have a non-empty intersection with the support of the continuous angular scattering function, i.e. with $\Xgam$. Therefore we can use the estimated UL support set to estimate $\Xgam$ as $\Xgamhat = \underset{i\in \suppulhat}{\Bcup} \Ic_{{\rm ul}, i}$.
This gives a fine approximation of the support of the angular scattering function, particularly when it has a block-sparse structure and $M\gg 1$. Now, the DL support is estimated by
\begin{equation}\label{eq:dl_supp_2}
\suppdlhat = \{ i\in [M] ~|~ \xidl\cap \Xgamhat \neq \emptyset \}.
\end{equation} 
In words, we determine those indices whose corresponding intervals intersect with the estimated support of the continuous angular scattering function and in this way obtain an estimate of the DL channel support set. This procedure takes place at the BS for all users, resulting in the example schematic of Fig. \ref{supp_sets}. 


\section{Sparsification, Probing and Estimation}\label{sec:4}
Using the estimated DL channel support information, in this section we propose a novel technique that allows a flexible channel probing depending on the amount of pilot overhead tolerated by the system. To estimate DL channels, the BS broadcasts $T$ pilot vectors in $T<N_c$ of its available signal dimensions in the same channel resource block to probe the channel vectors. Since the channel vector is a $M$-dimensional, to estimate the channel, in the conventional scheme the BS needs to probe it over at least $M$ signal dimensions, i.e. $T\ge M$. After collecting the $T \ge M$ pilot measurements, each user must feedback its estimated channel vector. The feedback can be done either using analog feedback, by sending the unquantized coefficients as I and Q symbols over the UL, or using quantized feedback, implemented by a variety of schemes \cite{jindal2006mimo}.

CS-based techniques exploit the sparsity of the channel vectors to reduce the necessary DL pilot dimension to $T=\Oc\left(s \log M\right)$, with $s$ being the sparsity order of the channel vector. However, these techniques also fail when the devoted resource for channel probing is less than the required amount. To address this issue, here we propose a new technique, referred to as active channel sparsification. The main idea is that since the angular support of all the users is estimated during UL pilot transmission, the BS can, depending on the resource budget devoted to channel training, design a pre-BF matrix such that the effective channel, which is the product of the pre-BF matrix and the actual physical channel, has a small dimension for every user. In this way, we can control the sparsity order of the DL channels and therefore obtain a flexible channel estimate.

We can formulate the problem of sparsification as follows. Given the estimated support sets, let $\Ac:=\cup_{k=1}^K  \suppdlhat^{(k)}$ denote the set of indices corresponding to all coupled angular directions. Now, we introduce the bipartite $\left|\Ac\right| \times K$ graph $\Gc=\left(\Ac,\Kc,\Ec\right)$ where on one side we have the elements of $\Ac$ and on the other side we have nodes, each corresponding to a support set $\suppdlhat^{(k)}$. An edge between element $a\in \Ac$ and set node $k\in \Kc$ exists if $a\in \suppdlhat^{(k)}$. Fig. \ref{bi_graph} illustrates $\Gc$ for the example estimated DL support sets in Fig. \ref{supp_sets}. Let $\bfW \in \{0,1\}^{|\Ac|\times K}$ be the adjacency matrix corresponding to the bipartite graph $\Gc$ and let $a$ denote the label of an angular direction. We know that $[\bfW]_{a,k}=1$ if and only if the angular direction $a$ is included in the set $\suppdlhat^{(k)}$. 
We want to maximize both the number of probed angular directions and the number of probed (and eventually served) users. Probing more angles is desirable because it increases the signal dimension and probing more users is desirable because then we can serve more users. In addition these two objectives are not in contrast. Now, we introduce two sets of binary variables $\{z_a\}_{a\in \Ac}$ and $\{u_k\}_{k\in [K]}$, where the first set represents the set of effective angular directions and the second set represents the set of users. The optimization problem can be formulated as follows,
\begin{equation}\label{eq:my_opt}
\begin{aligned}
& \underset{z_a,u_k \in \{0,1\} \forall a,k}{\text{maximize}} && \sum_{a\in \Ac} z_a  + \sum_{k\in [K]} u_k,    \\
& \text{subject to} && z_a \le \sum_{k\in [K]} [\bfW]_{a,k} u_k, ~ \forall a,  \\
& ~ &&  u_k \le \sum_{a\in \Ac} [\bfW]_{a,k}  z_a, ~ \forall k, \\
& ~ && \sum_{a\in \Ac} [\bfW]_{a,k} z_a \le M (1-u_k) + T, ~ \forall k.\\
\end{aligned}
\end{equation} 
Let us explain this problem in detail. The objective function in \eqref{eq:my_opt} stands for the total number of probed angular directions plus the number of served users. The first constraint ensures that if an angular direction $a$ is selected, i.e. if $z_a=1$, then there should exist at least one served user, e.g. user $j$ ($u_j=1$) that is coupled with this angular direction. The second constraint on the other hand ensures that if a user $k$ is served, i.e. if $u_k=1$, then there should exist at least one selected angular direction, e.g. angular direction $i$ ($z_i=1$) that is coupled with this user. The third constraint states that when when user $k$ is served, i.e. when $u_k=1$, the maximum number of selected angular directions coupled with this user is no more than $T$, satisfying the restriction to pilot dimension, while if $u_k=0$, this constraint is redundant. The problem \eqref{eq:my_opt} is an LIP optimization problem which can be easily solved by off-the-shelf solvers such as MATLAB. 

After obtaining the solution of problem \eqref{eq:my_opt}, i.e. $\left\{z_a^\ast \right\}_{a=1}^M$ and $\left\{u_k^\ast \right\}_{k=1}^K$, we obtain the set of probed angular directions as $\Bc=\{a:z_a^\ast=1\}$. Then we can define the pre-BF matrix $\bfB \in \bC^{|\Bc|\times M}$ by $\bfB = {\bf F}_{\Bc}^\herm$, where ${\bf F}_{\Bc}$ is a sub-matrix of the DFT matrix, formed by selecting those columns of $\bf F$ whose indices are in $\Bc$. Now, the effective DL channel coefficients vector $\heff \in \bC^{|\Bc|\times 1}$ is simply a product of the actual channel $\hdl$ and the pref-BF matrix $\bfB$, i.e. $\heff=\bfB \hdl$. For every user $k$, $\heff$ has less than $T$ non-zero elements. The locations of these elements are given by a set $\Omega_k\subset \{1,\ldots,|\Bc|\}$ and are known to the BS (each element in $\Omega_k$ has a one to one correspondence to an element in $\Bc \cap \suppdlhat^{(k)}$ which is the set of effective angular directions for user $k$). As a result, the BS can estimate the effective channel for all users by $T$ pilot vectors.

\subsection{Channel Probing and Estimation}
Following the previous discussions, if for a user $j$ we have that $u_j^\ast=0$, this user will not be probed and served because its effective dimension is larger than the pilot dimension. To estimate the effective DL channels belonging to other users, the BS broadcasts $T$ random Gaussian vectors in $T$ time-frequency signal dimensions. We denote the transmitted probing Gaussian vector in the $j$\textsuperscript{th} signal dimension by ${\boldsymbol \psi}_j \in \bC^{|\Bc|\times 1}$, and denote the set of BS probing vectors by a matrix ${\bf\Psi}  \in \bC^{T\times |\Bc|}$ where ${\bf\Psi}_{j,.}={\boldsymbol \psi}_j^\transp$. 
The received signal at user $k$ after pilot transmission is given by
\begin{equation}\label{eq:cs_eq_1}
{\bfy}^{(k)}= {\bf\Psi} \bfB {\check{\bfh}_{{\rm dl}}^{(k)}} + {\bfn}^{(k)} = {\bf\Psi} {\bfh^{(k)}_{\text{eff}}} + {\bfn}^{(k)},
\end{equation}
where ${\bfh^{(k)}_{\text{eff}}} $ is the effective DL channel coefficients vector for user $k$ and $\bfn^{(k)} \sim \cg({\bf 0},{\bf I}_T)$ is the AWGN vector at the user side, with i.i.d unit-variance entries. We assume that $\Vert {\bf \Psi}_{j,.}\Vert^2=\Pbf$ for all $j$, where $\Pbf$ is the power spent on a single probing vector by the BS. The measurement vector $\bfy^{(k)}$ is received at the BS for $k\in [K]$ via analog feedback. Then the BS uses its estimate of the DL support of each of the users to estimate their effective channel by
\begin{equation}
\left[\widehat{{\bfh}}_{\text{eff}}^{(k)}\right]_{\Omega_k}  = \left({\bf\Psi}_{.,\Omega_k} \right)^\dagger \bfy^{(k)},
\end{equation}
where $(\cdot)^\dagger$ is the Moore-Penrose pseudo-inverse, where $\left[\widehat{{\bfh}}_{\text{eff}}^{(k)}\right]_{\Omega_k }$ denotes the entries of $\widehat{{\bfh}}_{\text{eff}}^{(k)}$ whose indices are in $\Omega_k$ and ${\bf\Psi}_{.,\Omega_k}$ is a sub-matrix of ${\bf\Psi}$ formed by concatenating the columns whose indices are in $\Omega_k$. All other entries of $\widehat{{\bfh}}_{\text{eff}}^{(k)}$ whose indices are not included in $\Omega_k$ are set to zero.


For DL data transmission, we propose using the greedy zero-forcing (ZF) precoder to reduce inter-user interference. Having an estimate of the to-be-served DL channel vectors, this precoder selects a subset of linearly independent  vectors with maximum size. Without loss of generality we assume that the first $K'$ users are eventually served and derive the rate expression. 
%

Let $\widehat{\bfH}_{{\text{eff}}}=\bfB^\herm [ \widehat{{\bfh}}_{{\text{eff}}}^{(1)}, \ldots, \widehat{{\bfh}}_{{\text{eff}}}^{(K')}]$ be the matrix consisting of estimated effective DL channel vectors for the served users. The ZF precoder is a matrix denoted by ${\bfT}_{.,k} := \frac{\Qm_{.,k} }{\Vert \Qm_{.,k} \Vert},$ where $\Qm = (\widehat{\bfH}_{{\text{eff}}}^\herm)^\dagger$. The transmit signal at the BS is then given by $\bfx = \sqrt{\frac{P}{K'}}\bfT \bfs,$ where $\bfs \in \bC^{K'\times 1}$ is the vector of unit-power user symbols $s_k$ and $P$ is the  transmit power. The received signal at user $k$ can be written as
\begin{equation}
\begin{aligned}
r_k & = \sqrt{\frac{P}{K'}} (\check{\bfh}_{{\rm dl}}^{(k)})^\herm \bfT \bfs + n_k \\
& = \sqrt{\frac{P}{K'}}(\check{\bfh}_{{\rm dl}}^{(k)})^\herm \bfT_{.,k} s_k + \sqrt{\frac{P}{K'}}\underset{ j\neq k}{\sum}(\check{\bfh}_{{\rm dl}}^{(j)})^\herm \bfT_{.,j} s_j + n_k,
\end{aligned}
\end{equation} 
where $n_k \sim \cg(0,1)$. As a performance metric, we use the lower and upper rate bounds presented in \cite{DBLP:journals/corr/Caire17}. Define the variable $g_{k,k'} = \sqrt{\frac{P}{K'}}  (\check{\bfh}_{{\rm dl}}^{(k)})^\herm \bfT_{.,k}~,~k,k' \in[K'] $. For a user $k$, an upper and a lower bound for the rate are respectively given by
\begin{equation}\label{eq:rate_ub}
R_k^{\text{ub}} = ( 1 - \frac{T}{N_c}) \EE [ \log ( 1 + \frac{\left|g_{k,k'}\right|^2}{1 + \sum_{k'\neq k} \left|g_{k,k'}\right|^2}) ],
\end{equation}

\begin{equation}\label{eq:rate_lb}
\begin{aligned}
R_k^{\text{lb}} & = ( 1 - \frac{T}{N_c}) \EE [ \log ( 1 + \frac{\left|g_{k,k'}\right|^2}{1 + \sum_{k'\neq k} \left|g_{k,k'}\right|^2}) ]\\
& - ( 1 - \frac{T}{N_c}) \frac{1}{N_c} \sum_{k'=1}^{K'} \log \left(1+N_c\text{Var} \left( g_{k,k'}\right)\right)\\
\end{aligned},
\end{equation}
where $\text{Var}(\cdot)$ denotes the variance.

\section{Simulation Results}\label{sec:5}
In this section we provide numerical simulation results to assess the performance of our proposed algorithm empirically. We compare our algorithm with the CS based algorithm proposed in \cite{rao2014distributed}. This algorithm uses a joint orthogonal matching pursuit (J-OMP) method to reconstruct the sparse vector of channel coefficients $\hv_{{\rm dl}}^{(k)}$ from noisy measurements $\bfy^{(k)}$. The measurements in this method are obtained using a random Gaussian sensing matrix. We consider a BS with $M=128$ antennas and $K=20$ users. Adopting the GSCM channel model, we assume that there are three MPC clusters in the environment, located at random within the range $\Theta:=[-\theta_{\max}, \theta_{\max})$, each with an angular span $\approx \frac{2\theta_{\max}}{10}$, which is roughly equivalent to $\frac{M}{10}=\frac{128}{10}\simeq 13$ support elements. Each user is coupled with one, two, or three of these MPCs, chosen at random. Since we have three clusters in total, the maximum sparsity order of a channel vector is $s_{\max} = 3\times 13=39$. Unlike \cite{rao2014distributed}, we do not assume that the users share a common MPC, although this might be the case in a random setting. It is important to note that since we use a continuous scattering model, when the channel is represented in the Fourier basis, the corresponding vector of coefficients is not sparse in the strict sense, but rather well-approximated by a sparse vector. This is slightly different from the setting proposed in \cite{rao2014distributed}, where the channel is assumed to be strictly sparse. We feed the sparsity order of each channel vector to the J-OMP algorithm while this information is not provided to our proposed algorithm. We assume a resource block of size $N_c=128$ to be available at the BS. Table \ref{tab:table_1} summarizes the main parameters used in our simulations. 	
\begin{table}[t]
	\centering
	\begin{tabular}{ |l|c|c| }
		\hline
		\multicolumn{3}{ |c| }{Simulation Parameters} \\
		\hline
		Maximum Angular Range & $2\theta_{\max}$ & $\frac{2\pi}{3}$ \\ \hline
		Antenna Spacing & $d$ & $\frac{\lambda_{{\rm ul}}}{2 \sin(\theta_{\max})}$\\ \hline
		Carrier Frequency over DL band & $f_{{\rm dl}}$ & $\approx 1.1 f_{{\rm ul}}$ \\ \hline
		Number of Antennas & $M$ & $128$ \\ \hline
		Number of Users & $K$ & $20$ \\ \hline
		Number of Uplink Pilot Symbols & $L$ & $10$ \\ \hline
		Downlink Transmit Power & $P$ & $M\times \text{\small Downlink} ~\SNR$ \\ \hline
		Resource Block Size & $N_c$ & $128$ \\ \hline
	\end{tabular}
	\caption{Table of simulation parameters.}
	\label{tab:table_1} 
\end{table}

During the simulations the users transmit their UL pilots each over $L=10$ orthogonal dimensions. The SNR for UL transmission is set to $15$ dB. Then, the BS estimates the UL angular support of each user according to the method described in section \ref{sec:3} and interpolates the DL support set. Note that this process takes place only once for a fixed geometry and the information is used for all instantaneous channel realizations. If the pilot dimension $T$ is not less than the maximum user support size, i.e. $T\ge \underset{k}{\max}\left|  \suppdlhat^{(k)} \right|$, active channel sparsification is trivial and one can assume the set of probed angular directions to be the union of effective beam directions among all users. However, when $T< \underset{k}{\max}\left|  \suppdlhat^{(k)} \right|$, active channel sparsification is crucial, otherwise channel estimation has a large error. We adopt the method developed in section \ref{sec:4} to perform active channel sparsification. Obviously this step is not performed for the J-OMP method. After estimating DL channels, we construct the Greedy ZF precoding matrix for our proposed method and the J-OMP method. The ZF precoding matrices are then used to transmit in the DL. 

\begin{figure}[t]
	\centering
	\includegraphics[width=0.46\textwidth]{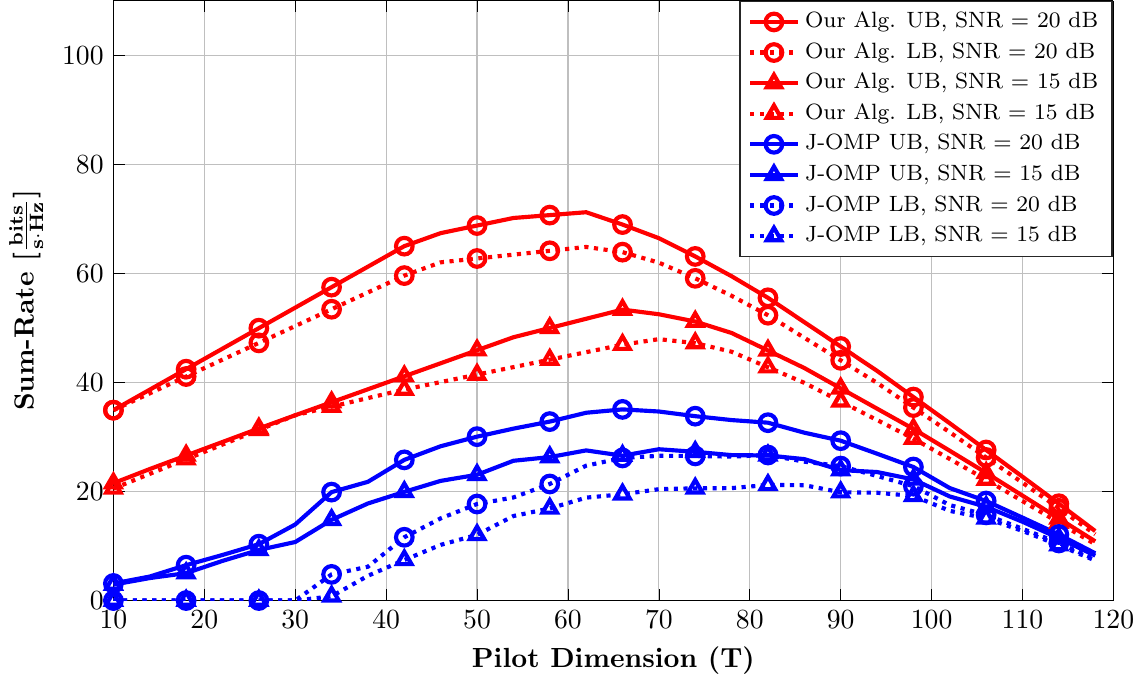}
	\caption{Comparison of Sum-Rate upper and lower bounds vs. pilot dimension $T$ for our algorithm and J-OMP with two different DL SNR Values.}
	\label{sweep_T}
\end{figure}

As a performance metric, we consider the sum-rate as a function of the available pilot dimension $T$ for two different values of Downlink SNR. To have a reliable channel estimate the J-OMP method needs a pilot dimension much larger than $s_{\max}=39$. However, using the estimated DL channel support information, our algorithm is capable of a highly accurate DL channel estimation using only $T\approx s_{\max}$ pilots. In addition, even for lower pilot dimensions, we obtain a good estimate of the channel vectors, using the active channel sparsification method. Fig. \ref{sweep_T} illustrates the comparison between our proposed method and the J-OMP method in terms of sum-rate lower and upper bounds formulated in \eqref{eq:rate_lb} and \eqref{eq:rate_ub}, calculated via Monte-Carlo simulations. This figure shows that our method achieves a much better performance compared with J-OMP, even with very low pilot dimensions. In fact the achievable lower bound in our method is higher than the rate upper bound of the J-OMP method for both SNR values.

\section{Conclusion}\label{sec:6}
In this paper, we presented a method for an efficient implementation of FDD massive MIMO systems, based on the idea of UL-DL reciprocity of the angular scattering function. Using this method dramatically reduces both the necessary DL pilot dimension and the feedback overhead, even compared with the state-of-the-art CS-based techniques. We further proposed the active channel sparsification method, which designs a pre-BF matrix to smartly reduce the effective channel dimension for each user, such that all channel vectors can be estimated with controllable error proportional to the available pilot dimension at the BS. Our simulation results show that our proposed method outperforms the state-of-the-art work based on CS in terms of achievable sum-rate.   

\balance
{\small
\bibliographystyle{IEEEtran}
\bibliography{references}
}

\end{document}